\documentclass[conference]{IEEEtran}
\IEEEoverridecommandlockouts
\usepackage{cite}
\usepackage{amsmath,amssymb,amsfonts}
\usepackage{algorithmic}
\usepackage{graphicx}
\usepackage{textcomp}
\usepackage{xcolor}
\usepackage{epstopdf}
\usepackage{amsmath,bm}
\def\BibTeX{{\rm B\kern-.05em{\sc i\kern-.025em b}\kern-.08em
    T\kern-.1667em\lower.7ex\hbox{E}\kern-.125emX}}

\usepackage[papersize={8.5in,11in}, left=0.625in, right= 0.625in, top=0.75in, bottom=1in]{geometry}

\begin{document}

\title{Mask R-CNN Based Object Detection for Intelligent Wireless Power Transfer \\
{\footnotesize}
}

%\author{\IEEEauthorblockN{Aozhou Wu}
%\and
%\IEEEauthorblockN{Qingqing Zhang}
%\and
%\IEEEauthorblockN{Wen Fang}
%\and
%\IEEEauthorblockN{Hao Deng}
%\and
%\IEEEauthorblockN{Sai Jiang}
%\and
%\IEEEauthorblockN{Qingwen Liu}
%
%\IEEEauthorblockN{
%  \textit{Department of Computer Science and Technology} \\
%    \textit{Tongji university}\\
%    Shanghai, People's Republic of China \\
%    1732968@tongji.edu.cn}
%}

\author{
\IEEEauthorblockN{Aozhou Wu, Qingqing Zhang, Wen Fang, Hao Deng, Sai Jiang, Qingwen Liu}
\IEEEauthorblockA{\textit{Department of Computer Science and Technology} \\
\textit{Tongji University}\\
Shanghai, People's Republic of China \\
\{1732968, anne, wen.fang, denghao1984, sai.jiang, qliu\}@tongji.edu.cn}
}

\maketitle

\begin{abstract}

Resonant Beam Charging (RBC) is a promising multi-Watt and multi-meter wireless power transfer method with safety, mobility and simultaneously-charging capability. However, RBC system operation relies on information availability including power receiver location, class label and the receiver number. Since smartphone is the most widely-used mobile device, we propose a Mask R-CNN based smartphone detection model in the RBC system. Experiments illustrate that our model reduces the smartphone scanning time to one third. Thus, this machine learning detection approach provides an intelligent way to improve the user experience in wireless power transfer for mobile and Internet of Things (IoT) devices.

\end{abstract}


\section{Introduction}

\begin{figure}[htbp]
\centering{\includegraphics[scale = 0.19]{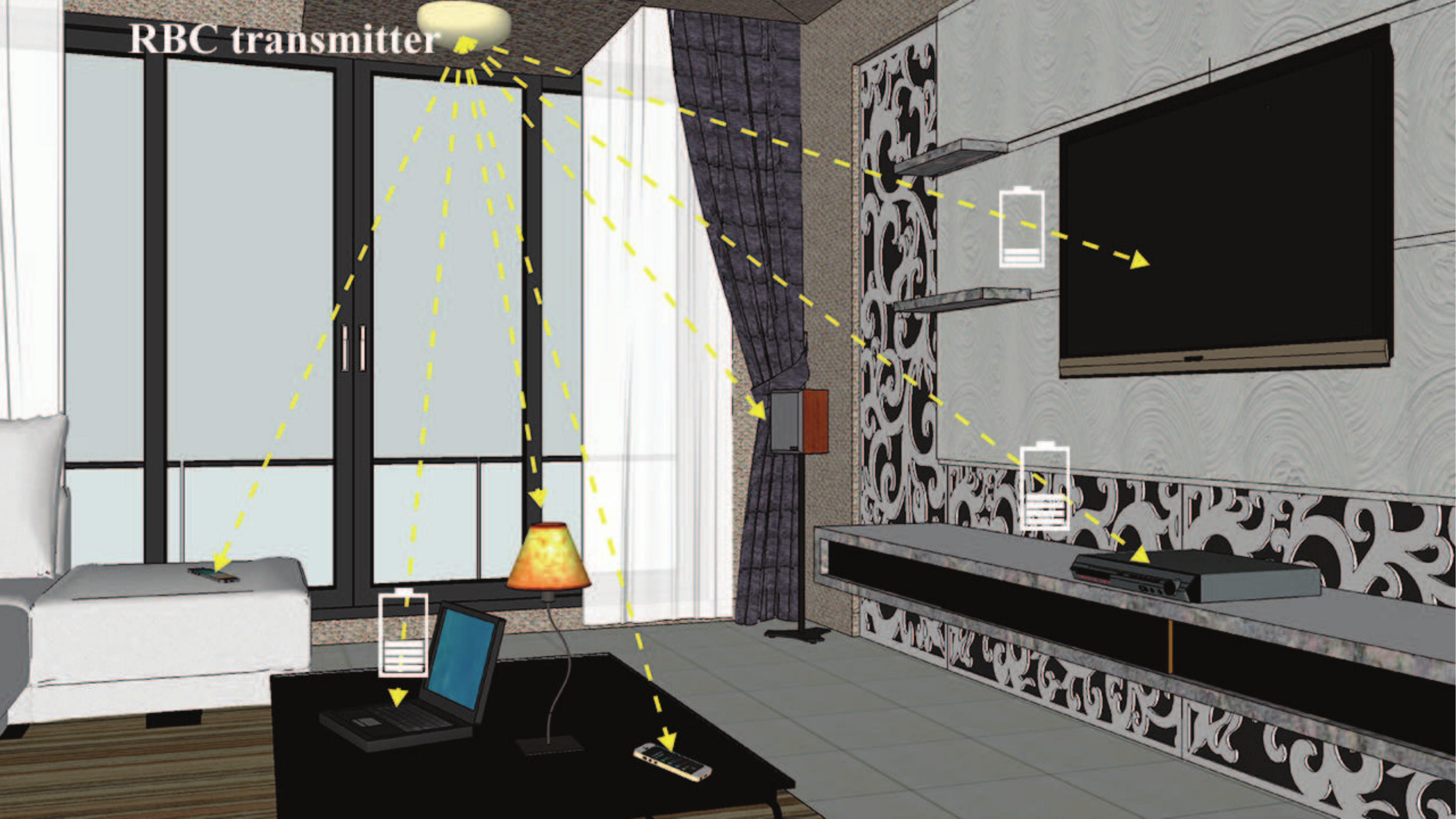}}
\caption{An application scenario with RBC system. Multi-device, e.g., mobile phone, laptop, TV, lamp and so on, can be charged simultaneously.}
\label{RBC}
\end{figure}

In recent years, with the rapidly increasing contradiction between battery endurance and power supply of electronic devices, wireless charging techniques are getting more and more attention. However, traditional solutions, e.g., inductive coupling, magnetic resonance coupling, radio frequency and so on, face the challenges of distance, power, safety, and mobility. An effective approach of wireless charging is Resonant Beam Charging (RBC) presented in \cite{rbc}. In the RBC system, multi-device can be charged with Watt-level power at meter-level distance, similar to Wi-Fi. Moreover, when the line of sight between the RBC transmitter and receiver is blocked by any object, wireless power transfer will stop immediately. Thus, the RBC system can deliver power safely. Therefore, electronic devices can be charged anywhere, anytime within the coverage of the RBC system. Fig.~\ref{RBC} presents an application scenario of the RBC system.

To charge devices intelligently, the certain information of the RBC receivers, including receiver location, class label and the receiver number, are necessary for the RBC transmitter. The location information provides the transmitter with a reference of orientation to transmit power. Judging by the class label, the transmitter can decide the corresponding charging operation. According to the receiver number, the transmitter can schedule the transmitting power more reasonably for charging receiver more efficiently. The receiver number can be worked out by summing class labels up. However, it is hard to obtain the class label and location information.

A traditional method to obtain the RBC receiver information is depicted as follows. Before charging, the transmitter will identify and locate the receiver firstly. The operation steps are: 1) The transmitter coverage area (decided by the field of view and the charging distance) is divided into $N$ small areas. 2) The transmitter scans each small area by sending searching signals. 3) If there are receivers in the area $N_i$, the transmitter will receive a feedback signal from the area $N_i$. 4) The feedback signal includes the accurate location information and class label of the receiver. However, the traditional method presented above is inefficient as the coverage area must be scanned one by one, and  sometimes the coverage area must be scanned entirely to identify and locate the receiver.

To improve the efficiency of obtaining receiver information, the area which is more likely to contain the receivers should be scanned at first. In other words, we find some candidate areas in the coverage area firstly and then scanning. This is similar to the object detection which detects location and class labels for some target objects in digital images. Building an object detection model to detect the RBC receivers, we can find some candidate areas with the output of detection model. Therefore, we present a method of charging object identification and positioning based on the object detection for the RBC system in this paper.

The contributions of this paper include: 1) We propose a Mask R-CNN based smartphone detection model to identify and locate the charging receiver for the RBC system using our own training dataset. 2) Our experiments illustrate that the detection model can reduce the identification and positioning time of the traditional method to one third.

In the rest of this paper, we will elaborate on the object detection in RBC system in section II. Then, we will build our object detection model in section III. In addition, we will show our experiments in section IV. Finally, we will give our conclusions and discuss open issues for future researches in section V.

\section{RBC Receiver Detection}

In this section, we present how to apply object detection to RBC system. Then we choose an appropriate object detection framework for our detection.

\subsection{RBC Receiver Detection}

\begin{figure}[b]
\centering{\includegraphics[scale = 0.19]{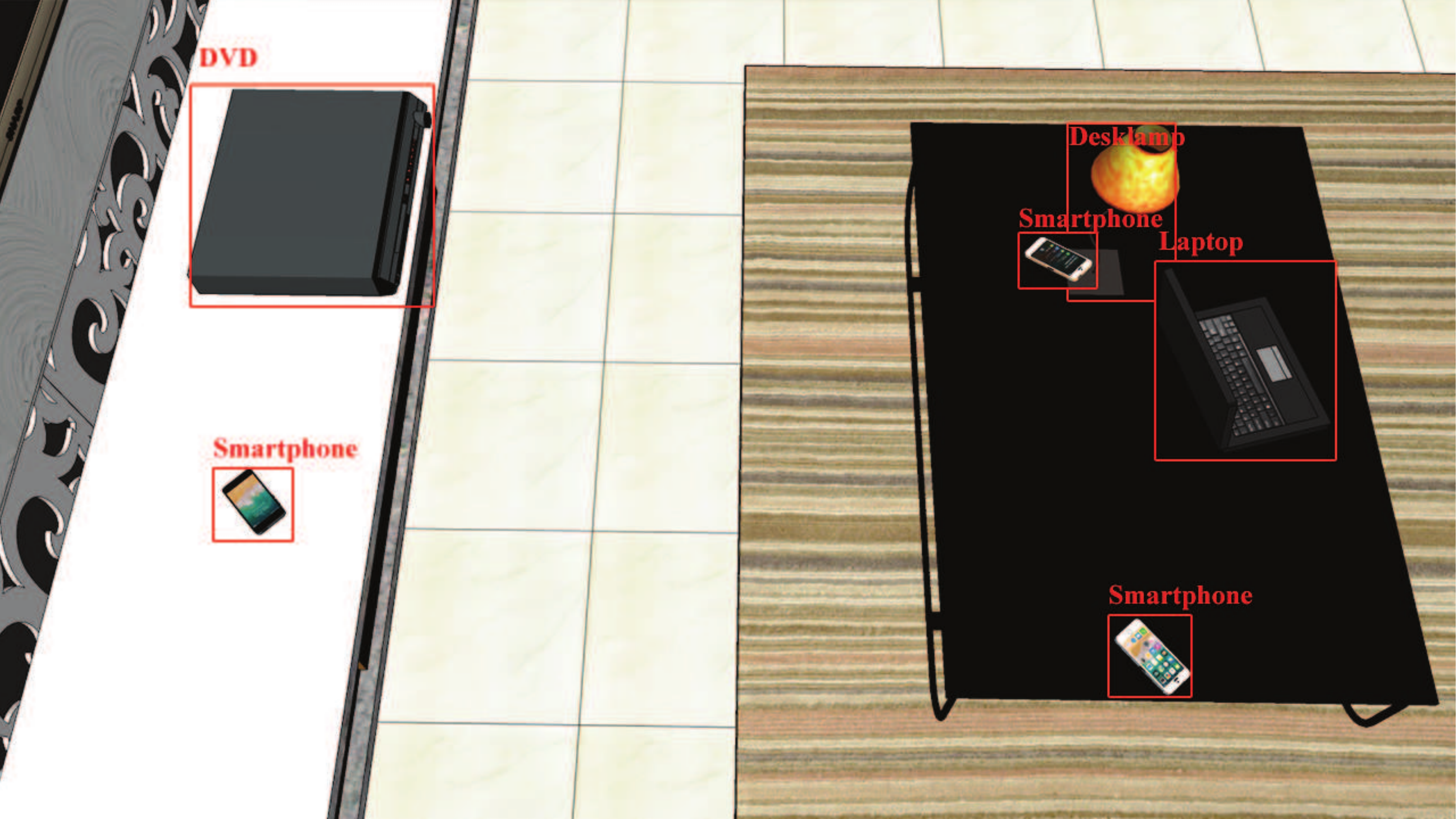}}
\caption{ Object detection for RBC system. The location information and class labels about the RBC receivers are extracted from the digital image}
\label{roomdetec}
\end{figure}

To combine the RBC system with object detection, we integrate a camera into the RBC transmitter to photograph the coverage area. We build the object detection function into the transmitter system to extract the location information and class labels about the receivers from these photos. Fig.~\ref{roomdetec} presents an example of detection result.

After detection, we calculate the central point with the location information. We may have several central points. The area where the central point lies in becomes the candidate area. Then, the transmitter will scan the candidate area. If the detection result is right, we will find the receiver at the first scanning. This is why our detection model makes the charging procedure more efficient. If wrong, we will scan the other areas after scanning candidate areas.

\subsection{Detection Frameworks}

Nowadays, Convolutional Neural Network (CNN) has become the main tool for object detection. There are two types of object detection frameworks based on CNN: One-stage framework and Two-stage framework. One-stage framework, which includes YOLOv3, SSD, is fast but with low accuracy\cite{yolo3,ssd}. Two-stage framework, such as Faster R-CNN and Mask R-CNN, is slow but with higher accuracy \cite{fasterrcnn,maskrcnn}. Considering the real application scenarios, the following questions should be taken into account: 1) We should pay more attention to small objects (smaller than 32piexl $\times$ 32piexl), since the size of targets in image may be very small like shown in Fig.~\ref{roomdetec}. 2) Detection within shorter time is better, because we have to consider the user experience.

Table I presents a comparison among the four frameworks with the COCO dataset in \cite{yolo3,maskrcnn} on an Nvidia Tesla M40 GPU. We report the standard COCO metrics including mean Average Precision(mAP), Average Precision for small objects (AP$_S$) and the detection time of an image \cite{coco}.

%To calculated mAP, Average Precision (AP) at 10 Intersection over Union (IoU) thresholds (from 0.50 to 0.95 step by 0.05) should be tested firstly. Then, mAP is calculated as the average value of AP. If an object in image is smaller than 32pixel$\times$32pixel, it will be considered as a small object. AP$_S$ is the mAP tested in small objects.

\begin{table*}[htbp]
\caption{Comparison among four main frameworks}
\begin{center}
\begin{tabular}{c|c|ccc}
\hline
& \textbf{backbone}&\textbf{mAP(\%)}&\textbf{AP$\bm{_S}$(\%)}&\textbf{time(ms)} \\
\hline
Faster R-CNN& ResNet-101-FPN& 36.2& 18.2& 175\\
%\hline
Mask R-CNN& ResNet-101-FPN& 38.2& 20.1& 195 \\
%\hline
SSD513& ResNet-101& 31.2& 10.2& 125  \\
%\hline
YOLOv3& Darknet-53& 33.0& 18.3& 51  \\
\hline
%\multicolumn{5}{l}{Object detection single-model results on COCO test-dev.}
\end{tabular}
\label{tab1}
\end{center}
\end{table*}

Mask R-CNN costs about 0.2s per image for detection and performs better than the other three frameworks according to the value of mAP and AP$_S$. Though YOLOv3 runs faster, we choose Mask R-CNN because its detection time as 0.2s is acceptable to us. Due to access certification and power allocation, the RBC system actually needs about 2s to get ready before delivering power. After weighting accuracy and speed, We build our RBC receiver detection model based on the Mask R-CNN framework.

\section{RBC Receiver Detection Implementation}

This section is about the implementation details for building our receiver detection model. We need a dataset to train our model. The RBC receiver can be integrated in various electronics, e.g., smartphone, laptop, Internet of Things (IoT) device and so on. The overall receiver dataset is huge and hard to establish. The smartphone ,as the most commonly used electronic device, is considered as the detection object in the RBC receiver detection model in this paper.

We build the detection model based on Mask R-CNN framework under the guidance of \cite{maskrcnn} at first. Then, we make a dataset for smartphone detection. Training is the final step which needs patience to tune and optimize. Here are the details of these three steps.

\subsection{Mask R-CNN Architecture}

\begin{figure}[b]
\centering{\includegraphics[scale = 0.26]{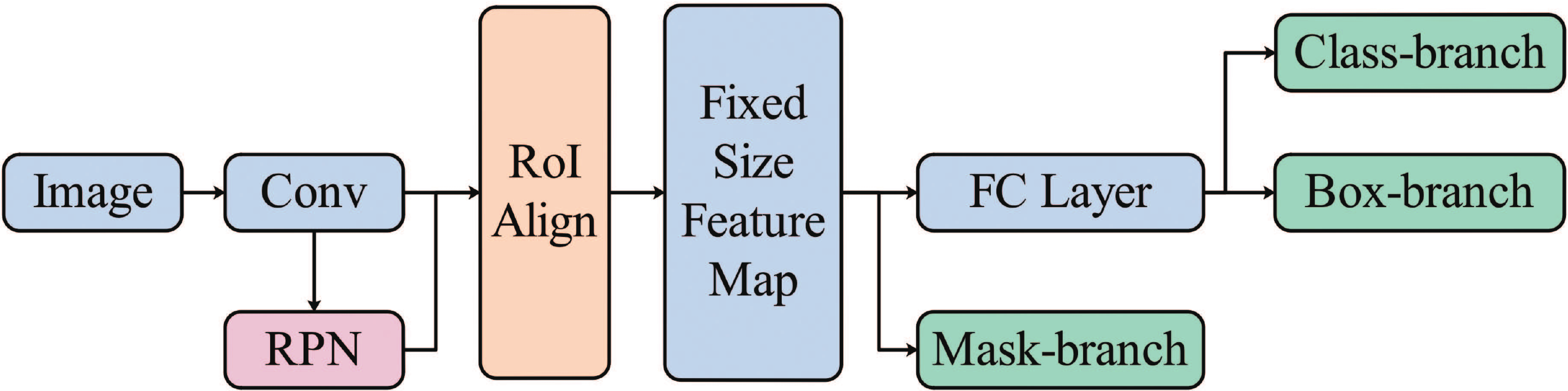}}
\caption{ Mask R-CNN architecture. Taking an image as input, the Mask R-CNN model will output class labels, bounding box coordinates and object masks of several objects.}
\label{maskrcnn}
\end{figure}

Fig.~\ref{maskrcnn} depicts the network architecture of Mask R-CNN. The Mask-branch may be superfluous at first. Since we aim at object detection, masks are not needed. But actually the Mask-branch increases the accuracy of object detection owing to multi-task learning \cite{maskrcnn}. So our network structure retains the Mask-branch.

The final layer of the Class-branch is a softmax layer, which is often used for multi-target classification. Though we only detect smartphones this time, the targets of wireless charging also include laptops, table lamps and IoT devices. So this softmax layer is also retained.

RPN layer was first proposed in \cite{fastrcnn}. It outputs several candidate regions on the original image by calculating feature maps. Combining feature maps with RPN layer, the RoIAlign layer obtains multiple Region of Interest (RoI) with a fixed size. The RoIAlign layer solves the problem of regional mismatch caused by the quantitative operation with the bilinear interpolation method \cite{b8}.

We use ResNet networks of depth 101 layers to extract features from images \cite{resnet}. With convolution and pooling operations, higher-level information is becoming more abundant, and the reserved space information is becoming less. This is why it is difficult to detect small targets if we only extract features from the final convolutional layer. Therefore, Feature Pyramid Network (FPN) is introduced which was proposed in \cite{fpn}. FPN uses a top-down architecture to build a feature pyramid with the last few convolutional layers. The feature maps of each layer now have appropriate advanced features with spatial information.

\subsection{Dataset Building}

Our dataset includes 1,600 original images. Since RBC system is applied in real indoor scenes, images are photographed from different places including classroom, office, dormitory, laboratory, etc. We use 8 different smartphones to represent different types of smartphones, so that our model is able to detect all kinds of smartphones. Moreover, smartphones are photographed in different situations, such as facing the lens of the camera directly and siding to lens. We pay particular attention to photographing at different distance to simulate real application scenarios.

In deep learning, richer data usually makes better results. So, we make an axisymmetric transformation to double our dataset. Finally, we add marks manually for each image with class labels and object masks (bounding boxes equal to the minimum circumscribed rectangle of the object masks). In the end, we get 3,200 data images and divide these images into the training set with 1,600 images, the development set with 800 images and the test set with 800 images.

\subsection{Model Training}
Since the number of our dataset is small, we initialize our network parameters with a COCO-pre-trained model. It means the weight parameters have already been trained with tens of thousands of images for several days at the time of initialization. Our training work is actually a fine-tuning process to make the output more targeted to smartphone detection.

\begin{figure}[b]
\centering{\includegraphics[scale = 0.19]{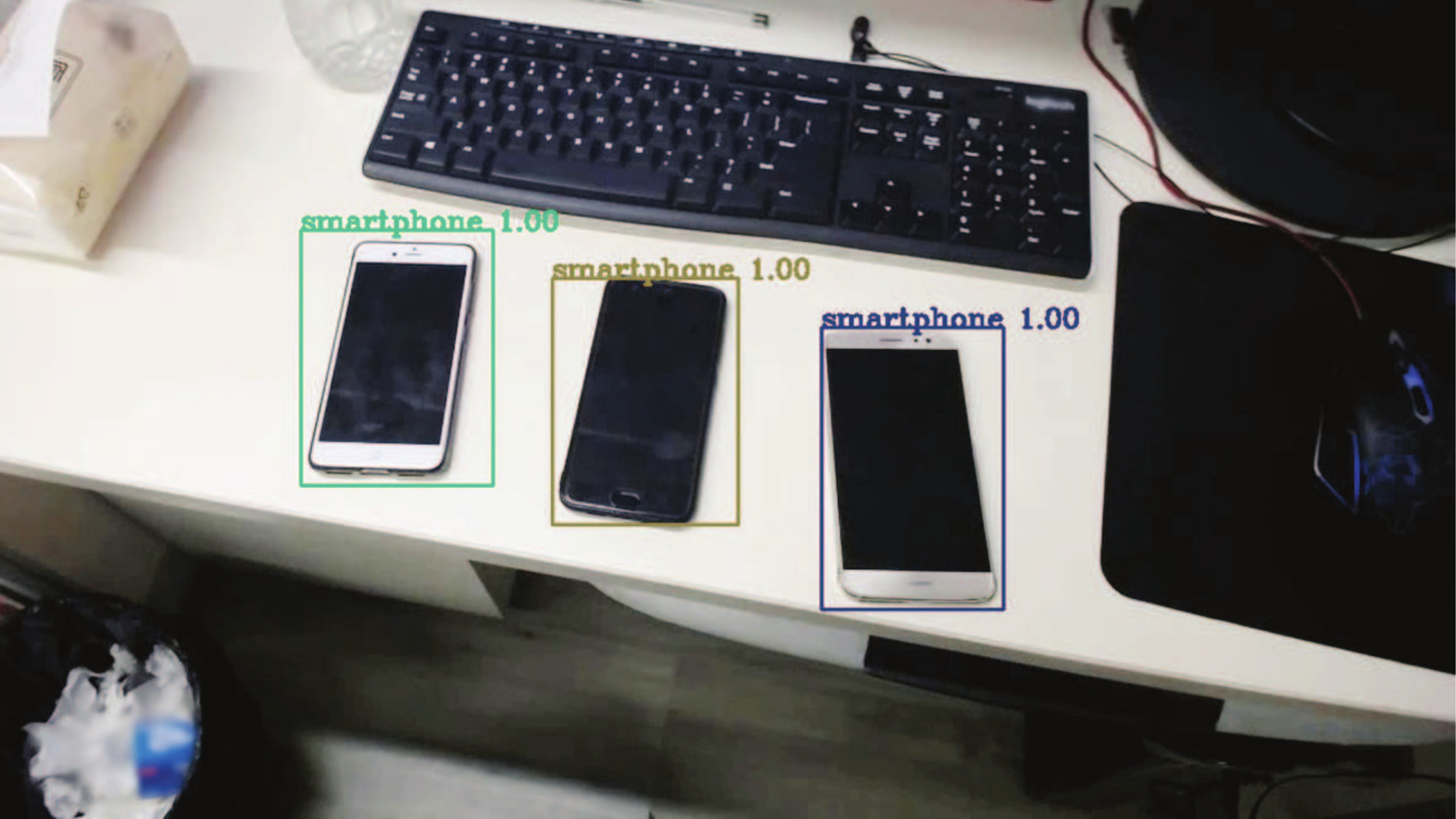}}
\caption{ Smartphone detection for RBC system. Each smartphone is marked with a bounding box and a class label.}
\label{deteg}
\end{figure}

The training process is divided into two steps. First, fine-tune the last three branches of the network model to enhance the feature extraction ability of smartphones. Then, use a smaller value of learning rate to fine-tune all layers. Under-fitting and over-fitting should be prevented both in the training set and development set. Fig.~\ref{deteg} shows the detection result of our model.

\section{Experiment Analysis}
In our experiments, we use Average Precision (AP) to evaluate our model. Our test set includes 800 images.
First, Our model is tested on this test set over different Intersection over Union (IoU) thresholds. In the test set, there are 400 special images which are photographed in different distance. The second experiment tests AP over different distance is based on these 400 images. We combine RBC system with our smartphone detection model in the third experiment.

\subsection{AP over Different IoU Threshold}

\begin{figure}[t]
\centering{\includegraphics[scale = 0.6]{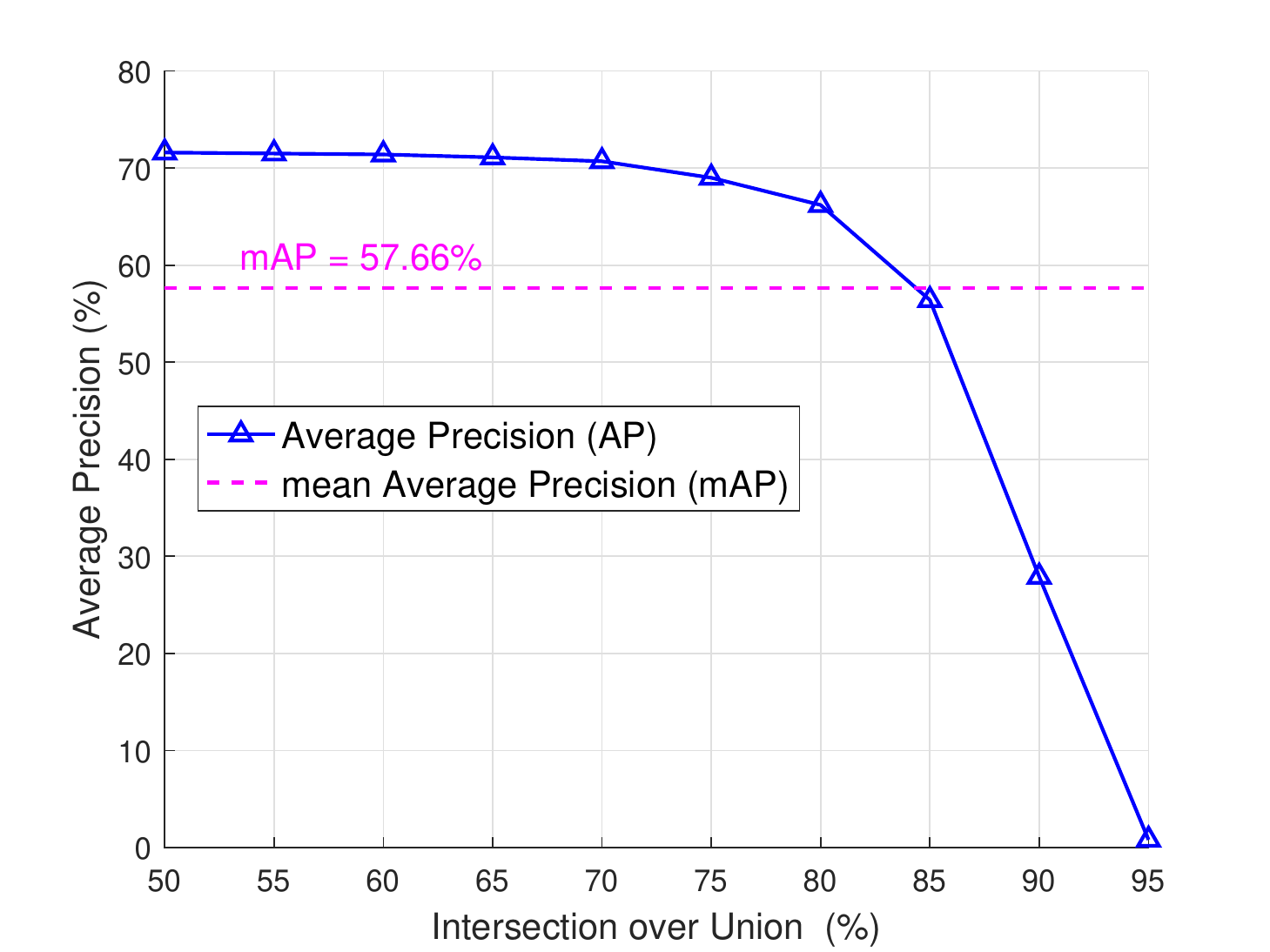}}
\caption{ Average Precision vs. Intersection over Union.}
\label{apiou}
\end{figure}

We test the AP value of our model at 10 IoU thresholds (from 0.50 to 0.95 step by 0.05) and calculate the average value mAP in Fig.~\ref{apiou}. Higher IoU means our output location is more closer to our object. Higher AP means our detection result is more accurate. The mAP 57.66 is much higher when comparing to mAP 38.2 which was tested on COCO test-dev \cite{maskrcnn}.

There are three reasons which cause high AP value: 1) Our classification task is simple, since we only classify smartphone and background. If we need multi-task detection, our AP may get lower.
2) Our test set contains many large targets which means the size of the smartphone in the image is large. Since detecting large objects is easier, the AP value becomes higher after averaging. We need more images with small targets to enrich our dataset.
3) Our scenarios are a bit monotonous, mainly concentrated in the laboratory, dormitory and classroom. The AP value goes up since there are many scenes overlapped between training set and test set. We need more images from various scenes both for the training set and the test set. Despite these factors, our experimental results are still good.

\subsection{AP over Different Distance}

\begin{figure}[t]
\centering{\includegraphics[scale = 0.6]{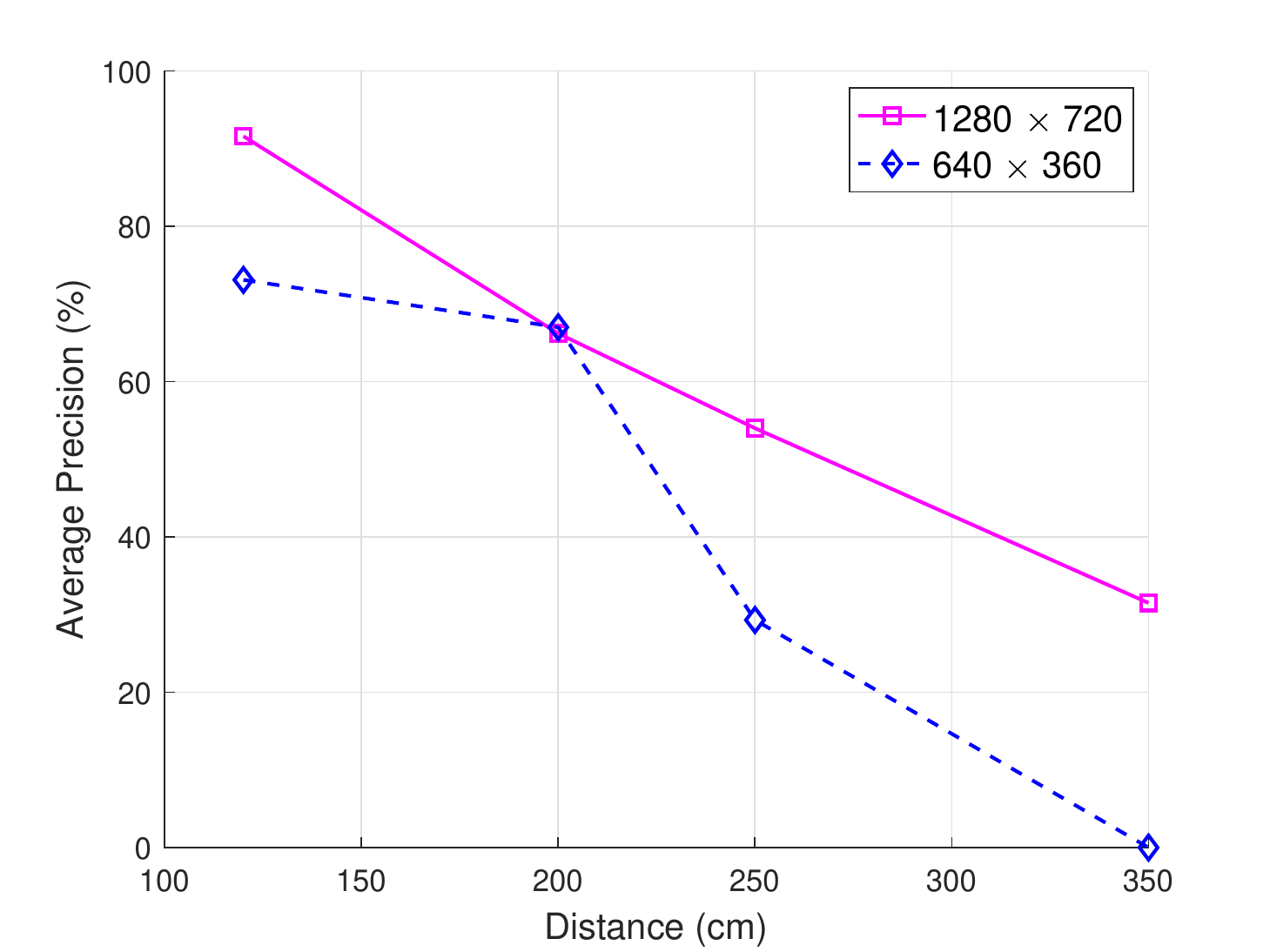}}
\caption{ Average Precision vs. Distance.}
\label{aparea}
\end{figure}

The detection performance in different distance is critical. In real application scenes, the distance between smartphones and the RBC transmitter always change. In consideration of the transmitter is usually installed on the ceiling, we take photos at 120cm, 200cm, 250cm and 350cm, each with 100 photos. We save our photos in two different sizes: 1280pixel $\times$ 720pixel and 640pixel $\times$ 360pixel. We test AP at IoU 0.50. The detection results are given in Fig.~\ref{aparea}.

Since the distance is not a direct influence factor, the AP value varies with the image size if with same distance. The direct factor is the object size in image. For example, we take a photo for smartphone, 14$\times$7cm$^2$, which facing the lens directly from 120cm away. Then, we save the image in the size of 1280$\times$720. The smartphone size in that image is 124pixel $\times$ 62pixel which is large. If the image is saved in the size of 640$\times$360, the smartphone size will reduce to 62$\times$31, which is in middle size. Object detection model performs better in large objects. This is the reason why our smartphone detection model performs better in image 1280$\times$720 than image 640$\times$360.

There is a special case at the distance of 200cm where the AP of image 640$\times$360 is a little higher. This is caused by some objects such as computer screen and mouse pad are detected as a smartphone in the image 1280$\times$720. With more detected smartphones at the same time, more other irrelevant items are misidentified. This makes the value of AP finally reduced.

In our experimental environment, when the distance is about 350cm, our smartphone size is about 28$\times$14 in image 1280$\times$720 with AP 31.5. As for image 640$\times$360, the smartphone size in image is too small to detect. Therefore, we should ensure that our object size in image is larger than 30$\times$15 ,or it will be undetectable. We may use our model within 3 meters in our experimental environment.

\subsection{Benefits on RBC System}
Combining smartphone detection model with the RBC system reduces the time of locating and identifying receivers. In the traditional method, we divide the coverage area into $N$ small areas before charging. Time for scanning each area is $T_s$. We need to scan each small area one by one. The average scanning time $T_1$ is depicted as:

\begin{equation}\label{chargetime-capacity}
{T}_{1}= \frac{(1+N) T_{s}}{2}.
\end{equation}

Our detection model takes $T_d$ to detect smartphones. We start scanning from the candidate area at first. The probability of the candidate area including smartphones equals to $AP$. If we do not find a smartphone at the first scanning, other areas will be scanned. The average scanning time $T_2$ is depicted as:

\begin{equation}\label{chargetime-capacity}
{T}_{2}= T_{d} + AP T_{s} + (1-AP)(1 + \frac{N}{2}) T_{s}.
\end{equation}

\begin{figure}[t]
\centering{\includegraphics[scale = 0.6]{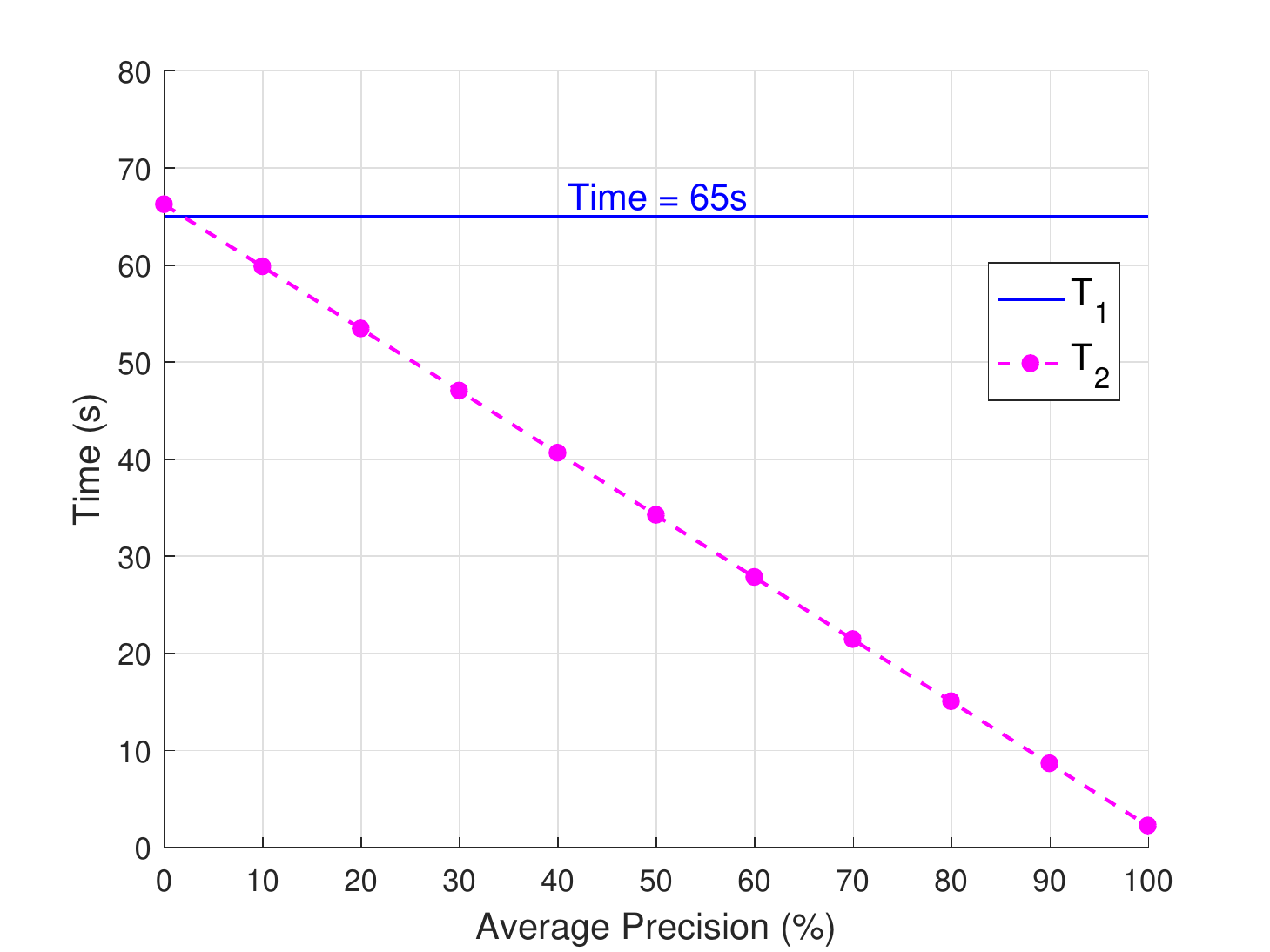}}
\caption{ Scanning Time vs. Average Precision.}
\label{timeap}
\end{figure}

In RBC system, we usually make $N$ equals to 64. Each area costs 2 seconds to scan \cite{rbc}. Our detection model costs 0.2 seconds per image. Fig.~\ref{timeap} presents the association between $T_2$ and $AP$ and also marks $T_1$. In this situation, $T_1$ equals to 65 seconds. In our detection model, AP value is larger than 0.70 in the IoU of 0.50 according to Fig.~\ref{apiou}. Taking AP equals to 0.70 as an example, we get $T_2$ equals to 21.4 seconds. So, without detection model, we take 65 seconds to locate smartphones in average. With detection model, we only need 21.4 seconds in average, cutting the time down to one third.

\section{Conclusions}

In this paper, we adopt Mask R-CNN based object detection to make wireless charging more intelligent. We build a smartphone detection model to extract information including the smartphone location, class label and smartphone number for the RBC system. We train the detection model using our own smartphone dataset. Experiments show that our model cuts the RBC scanning time down to one third. This machine learning detection approach provides an intelligent way to improve the user experience in wireless power transfer for mobile and IoT devices.

Due to the space limitation, there are several important issues unaddressed in this paper which are deserved to be studied in future. For example:

\begin{itemize}
\item In deep learning, richer data usually contributes to better results. Our dataset can be extended in several ways, such as taking more photos in more scenarios, using different smartphones, photographing smartphones in more situations with both the front sides and the back sides, etc.
\item Since there are many types of wireless charging receivers, the target category should be expanded to detect more varieties.
\item We implemented our model on a PC-level GPU. When building the detection function into the transmitter system which may not have a GPU, optimization is necessary.
\item The development speed of object detection frameworks is always beyond our imagination. Though we used Mask R-CNN framework at this stage, we believe there will be a framework which can work better both in accuracy and speed.
\item The detection model performs better when the smartphone is exactly facing the lens of the camera. Integrating multi-camera into the RBC transmitter to capture images from different angles is worthy to research.
\end{itemize}

\end{document}